\documentclass[final,5p,twocolumn,sort&compress,super,authoryear]{elsarticle}

\usepackage[justification=centering,font=small]{caption}
\usepackage{float}
\usepackage{geometry}
\usepackage{natbib}
\usepackage{setspace}
\usepackage{xkeyval}

\usepackage{fancyhdr}
\makeatletter
\def\ps@pprintTitle{%
  \let\@oddhead\@empty
  \let\@evenhead\@empty
  \def\@oddfoot{\reset@font\slshape{Accepted for publication in The Astrophysical Journal}\hfil\slshape{25 July 2017}}
  \let\@evenfoot\@oddfoot
}
\makeatother

\usepackage[version=3]{mhchem} 

\usepackage[T1]{fontenc}       
\usepackage{tgschola}

\usepackage{titlesec}

\titleformat*{\section}{\large\bfseries}
\titleformat*{\subsection}{\normalsize\bfseries}

\usepackage{graphicx}
\usepackage[figuresright]{rotating}
\usepackage{booktabs}
\usepackage{footnote}
\usepackage[hang,splitrule]{footmisc}
\usepackage{amsmath,amsfonts,mathabx}
\usepackage{balance}

\makeatletter
\newcommand\footnoteref[1]{\protected@xdef\@thefnmark{\ref{#1}}\@footnotemark}
\makeatother

\makeatletter
\DeclareRobustCommand*{\bfseries}{%
  \not@math@alphabet\bfseries\mathbf
  \fontseries\bfdefault\selectfont
  \boldmath
}
\makeatother

\newcommand{\wn}{\ensuremath{\mathrm{cm}^{-1}}}

\usepackage[driverfallback=dvipdfm]{hyperref}
\usepackage{xcolor}
\hypersetup{
    colorlinks,
    linkcolor={red!50!black},
    citecolor={blue!50!black},
    urlcolor={blue!80!black}
}

\begin{document}

\begin{frontmatter}

\title{\ce{C60+} and the Diffuse Interstellar Bands:\\An Independent Laboratory Check}


\author{Steffen Spieler, Martin Kuhn, Johannes Postler, Malcolm Simpson, Roland Wester, and Paul Scheier}
\address{Institut f{\"u}r Ionenphysik und Angewandte Physik, Universit{\"a}t Innsbruck \\
Technikerstrasse 25, Innsbruck A-6020, Austria}

\author{W. Ubachs}
\address{Department of Physics and Astronomy, LaserLaB, Vrije Universiteit Amsterdam \\
De Boelelaan 1081, NL-1081 HV Amsterdam, The Netherlands}

\author{and}
\address{}

\author{Xavier Bacalla, Jordy Bouwman, and Harold Linnartz\corref{correspondence}}
\cortext[correspondence]{Corresponding author. Email: linnartz@strw.leidenuniv.nl}
\address{Sackler Laboratory for Astrophysics, Leiden Observatory, Leiden University \\
P.O. Box 9513,  NL-2300 RA Leiden, The Netherlands}


\renewcommand{\abstractname}{\Large ABSTRACT}
\begin{abstract}
In 2015, Campbell et al. (Nature 523, 322) presented spectroscopic laboratory gas phase data for the fullerene cation, \ce{C60+}, that coincide with reported astronomical spectra of two diffuse interstellar band (DIB) features at 9633 and 9578 \AA. In the following year additional laboratory spectra were linked to three other and weaker DIBs at 9428, 9366, and 9349 \AA. The laboratory data were obtained using wavelength-dependent photodissociation spectroscopy of small (up to three) \ce{He}-tagged \ce{C60+}$-$\ce{He_$n$} ion complexes, yielding rest wavelengths for the bare \ce{C60+} cation by correcting for the \ce{He}-induced wavelength shifts. Here we present an alternative approach to derive the rest wavelengths of the four most prominent \ce{C60+} absorption features, using high resolution laser dissociation spectroscopy of \ce{C60+} embedded in ultracold \ce{He} droplets. Accurate wavelengths of the bare fullerene cation are derived based on linear wavelength shifts recorded for \ce{He_$n$}\ce{C60+} species with $n$ up to 32. A careful analysis of all available data results in precise rest wavelengths (in air) for the four most prominent \ce{C60+} bands: 9631.9(1) \AA, 9576.7(1) \AA, 9427.5(1) \AA, and 9364.9(1) \AA. The corresponding band widths have been derived and the relative band intensity ratios are discussed.
\end{abstract}


\end{frontmatter}


\section*{Introduction}
\label{Introduction}
\vspace{-3pt}
In 1922, Heger reported for the first time broad features, interstellar in nature, in spectra of stars reddened by cosmic dust \citep{Heger1922}. Today, more than 400 of such diffuse interstellar bands (DIBs) have been observed, covering the UV/VIS and near-IR with strongly varying spectral appearances \citep{Herbig1995,Hobbs2008,EDIBLES}. Despite decades of dedicated spectroscopic research, it has not been possible to identify their carriers \citep[see e.g.][]{Herbig1995,Snow2006,Sarre2006,Cox2014a}. Several claims of DIB identifications, like for \ce{C7-} \citep{Tulej1998}, \ce{C3H2} \citep{Maier2011b}, \ce{C4H2+} \citep{Krelowski2010}, or benzene derivatives \citep{Ball2000} had to be withdrawn, after additional laboratory and/or observational studies became available \citep{McCall2001,McCall2002,Araki2012,Liszt2012,Krelowski2012,Maier2011a,Araki2004}. Recently, however, a set of in total five near-infrared DIBs was assigned to one specific carrier, the fullerene cation \ce{C60+} \citep{Campbell2015,Campbell2016a,Campbell2016b,Walker2015,Walker2016}. The idea of \ce{C60+} acting as a DIB carrier dates back to \citet{Foing1994}, who linked matrix data recorded by \citet{Fulara1993} to two isolated DIBs, observed around 960 nm. A direct comparison between laboratory and observational data was not possible, as wavelengths recorded in matrix isolation experiments typically shift with respect to their gas phase equivalents through solid state interactions. It took 20 years to record for a first time the corresponding gas phase transitions and to provide their rest wavelengths \citep{Campbell2015,Campbell2016b}. The method that was used is based on trapping \ce{He}-tagged ions in an ion trap and recording mass spectrometrically the wavelength-dependent dissociation pattern \citep{Chakrabarty2013}. This method is powerful, as it guarantees full mass selectivity. Other methods, like plasma jets in combination with direct absorption spectroscopy \citep{Motylewski1999} suffer from the multitude of different species that are formed, but have the advantage that the spectra reflect fully isolated conditions, i.e., a laboratory spectrum can be directly compared to an astronomical spectrum (once corrected for any velocity offsets) \citep[see e.g.][]{Motylewski2000}. In an ion-tagged experiment this does not apply. The size of the inherent wavelength shift depends on the strength of the intermolecular bond with the atom tag and this, in turn, depends on where the tagging takes place. In particular, ionic complexes have shown to exhibit large wavelength shifts due to the strength of the intermolecular interaction, i.e., in proton bound or charge-transfer complexes \citep{Bieske2000}. In the case of \ce{He}-tagging, shifts due to molecular interactions are typically small. The resulting shifts involved in the \ce{C60+}$-$\ce{He_$n$} ($n=1-3$) case as presented by \citet{Campbell2015,Campbell2016a} were concluded to be of the order of 0.7 \AA~per tagged \ce{He} atom. Based on an extrapolation through the available data points, i.e., correcting for the offset induced by the tagged \ce{He} atoms, this has resulted in rest wavelengths in air that have been reported in several studies and adapted in follow-up papers, yielding 9632.1, 9577.0, 9427.8, 9365.2, and 9348.4 \AA, with $\pm$~0.2~\AA~accuracy \citep{Campbell2016b}. We will refer to these features as Bands 1, 2, 3, 4, and 5, respectively.

DIBs located at the laboratory wavelengths found for bands 1 and 2 have been reported in many previous observational studies. Even though the latter band overlaps with a stellar \ce{Mg\ \textsc{ii}} line, complicating the interpretation of the equivalent width \citep{Galazutdinov2017, Walker2017}, these features are considered as established DIBs. However, this is not the case for three weaker, previously unreported bands that were identified by \citet{Walker2015,Walker2016} on the basis of the laboratory data presented in the Campbell et al. studies. The overall S/N of these three DIB detections is quite low, which is mainly due to a very high level of telluric water pollution in the wavelength domain involved. Moreover, in these studies not all DIBs are visible along every line-of-sight and the total number of line-of-sights is limited. This puts strong constraints on the data interpretation.

The current situation can be summarized in the following way. The laboratory rest wavelengths of pure \ce{C60+} are based on a mass selective method of complexed ions, extrapolating numbers on the basis of a rather small number of data points (\ce{C60+}$-$\ce{He_$n$}, $n=1-3$). Five bands have been reported. The two stronger bands overlap with known DIB features; astronomical matches with the three weaker bands have been claimed by one astronomical team. This claim has obtained support in the literature \citep{Omont2016,Ehrenfreund2015}, but also raised serious concerns \citep{Galazutdinov2017}. In the latter work, a lacking correlation of the integrated absorbance between the two stronger DIBs is reported with intensity ratios deviating strongly from the laboratory values. Moreover, it appeared not possible to reproduce the weaker DIBs in spectra obtained along a large number of different lines-of-sight. A very recent study by \citet{Walker2017}, following a different approach, concludes on constant 9632/9577 equivalent width ratios in agreement with the laboratory values. As stated before, spectral pollution, mainly due to atmospheric water, severely complicates the analysis of all these ground-based data around 950~nm. For this reason, in another very recent study, \citet{Cordiner2017} presented the first Hubble Space Telescope data (i.e., telluric-free) around 1~$\mu$m. Several new near-infrared DIBs were reported, but the chosen target star was found to be suboptimal for the DIBs linked to \ce{C60+}.

As DIBs have been posing a paradigm for nearly a century now, it is important that the \ce{C60+} claims are confirmed or disproved by independent investigations. In the past, coincidental overlaps between laboratory spectra and DIB features have been mistakenly interpreted as matches. In the already mentioned case of \ce{C7-}, as many as eight vibronic bands were found to overlap with reported DIBs, even with comparable equivalent width ratios, and still, years later, this overlap was found to be coincidental \citep{Tulej1998,McCall2001}. Statistically, the chance of spectral overlap between laboratory and astronomical data is high, particularly in the visible range given the large number of DIB features. For this reason laboratory data with accuracies comparable to or better than achievable in astronomical observations are needed. Here we present laboratory data to provide an independent experimental verification of wavelength positions, linewidths, and line intensity ratios. A different experimental approach is used, based on embedding \ce{C60+} ions in ultracold \ce{He} droplets. This method is well established and has been successfully used to study both spectroscopic and dynamical properties of molecules \citep[e.g.][]{Brauer2011,Zhang2014,Bierau2010,Filsinger2012}. It has several advantages: it is mass selective, ultra-sensitive, and the very low temperatures allow adding many more \ce{He} atoms to the \ce{C60+} core than possible in an ion trap configuration. In a recent paper, we have discussed the solvation behavior of \ce{C60+} and discussed \ce{He_$n$}\ce{C60+} clusters with sizes larger than $n = 100$ \citep{Kuhn2016b}, showing that this process is governed by `onion-shell-like' filling properties. Here we focus on the first solvation shell that is of direct relevance for determining the \ce{C60+} rest wavelengths. An extension and reanalysis of the \citet{Kuhn2016a,Kuhn2016b} data is presented, providing accurate values that can be used to compare with the astronomical data.

Details of the applied method are described in the next section. Subsequently, our results are presented and discussed. We summarize with a conclusion.

\section*{Experiment and experimental results} \label{sec:experiment}

Helium nanodroplets are formed via expansion of 99.9999\% pure \ce{He} at a pressure of 2.2~MPa and a temperature of 9.7~K through a 5~$\mu$m pinhole nozzle into an ultra-high vacuum. Evaporative cooling after condensation leads to a final temperature of 0.37~K of the droplets \citep{Toennies2004}. The resulting log-normal size distribution of the neutral droplets yields an average size of about 200,000 \ce{He} atoms. The helium nanodroplets are loaded with fullerenes by passing the neutral beam through a stagnant vapor. Fullerenes are brought into the gas phase by vaporizing \ce{C60} powder (purity 99.99\%, SES Research) in a high temperature oven (290~$^{\circ}$C) that is mounted in a pick-up chamber. The \ce{C60+} fullerene cations are formed through electron impact ionization (60~eV) of the doped \ce{He} droplets and a subsequent charge transfer from the initially formed \ce{He+} atom to a \ce{C60} molecule that has been embedded into the \ce{He} nanodroplet. Low-mass ions that are ejected from the ionized droplet are then detected by a high resolution reflectron time-of-flight (TOF) mass spectrometer. The high mass resolution of m/$\Delta$m~$\sim$~3000 allows to discriminate between clusters with nominally the same mass. After ionization and before TOF detection, the cluster ions are irradiated by the narrow band output of a cw Ti:Sa laser with 0.6~W power and a spectral resolution of 10~MHz, covering 700 to 1000~nm. In the case of a resonant excitation of a \ce{C60+} transition, the photo-absorption causes the ion-\ce{He_$n$} complex to heat-up, resulting in the loss of weakly bound \ce{He} atoms. It typically takes an energy less than 100~\wn\ to evaporate one \ce{He} atom, so the resonant absorption of a 1~$\mu$m photon results in a substantial mass loss. This process can be visualized mass spectrometrically as a wavelength-dependent depletion in the mass signal. Mass spectra are recorded simultaneously for many different cluster ions while the laser is tuned over the wavelength domain where the four stronger \ce{C60+} bands are located. This is illustrated in Figure 1 for band 1, also for larger $n$-values that are not of direct relevance to the present study. Clearly, different cluster sizes absorb at different and characteristic wavelengths, and the resulting bandwidths are likely determined by small energy differences that correspond with different isomeric structures. Further experimental details are available from \citet{Leidlmair2012}.

\begin{figure}[ht!]
  \centering
  \includegraphics[width=0.4\textwidth]{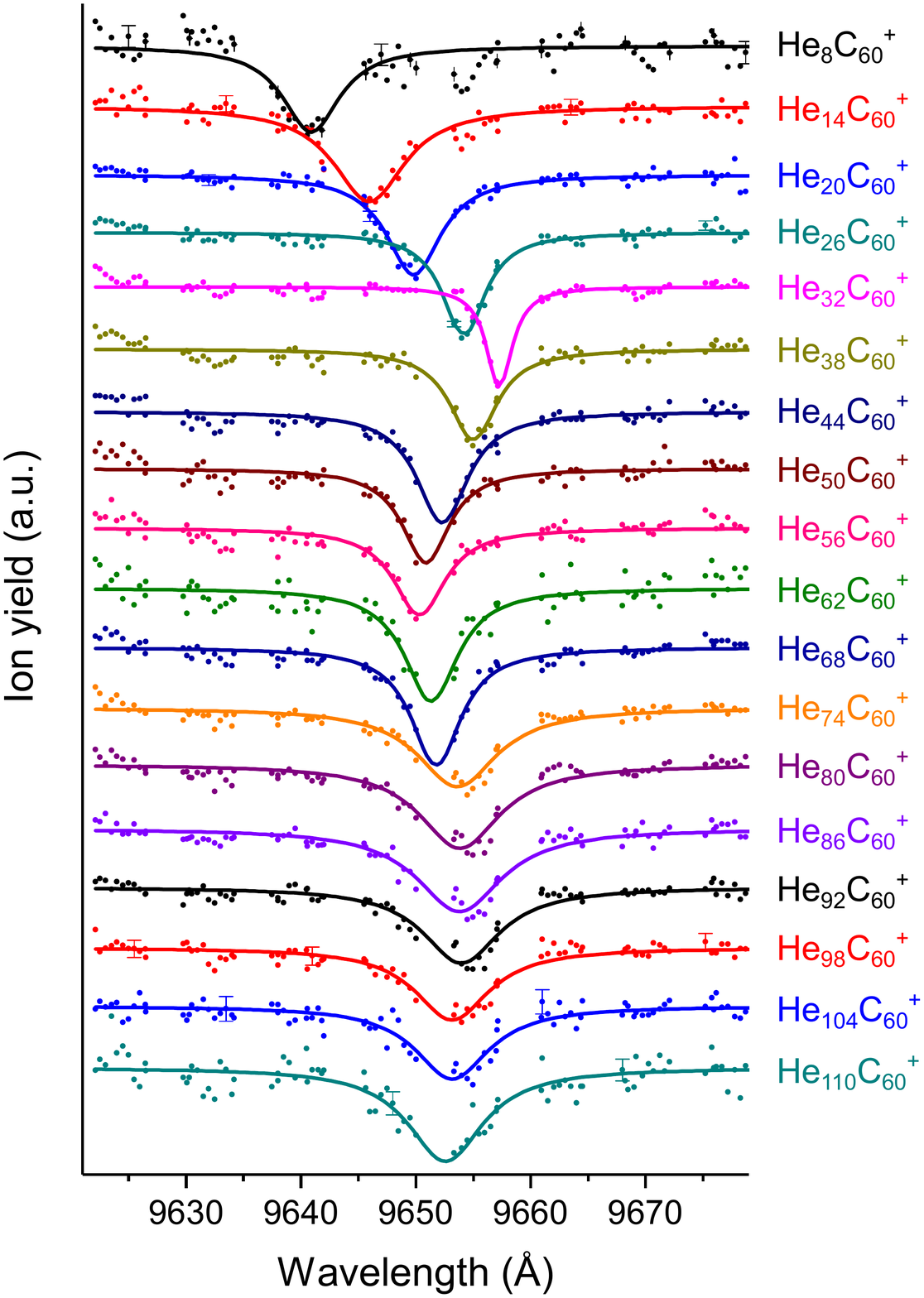}
  \caption{Ion yield of selected different cluster sizes as a function of the laser wavelength for band 1. Photoabsorption depletes the ion signal to a minimum at different wavelength positions. The resulting size dependent shifts for $n=8$, 14, 20, 28, and 32 (including all other $n<32$ values) are summarized in Figure~2. The wavelengths are given in vacuum, but at the resolution shown here this does not strongly deviate from air numbers. The smallest linewidth is recorded for \ce{He32C60+}.}
  \label{fig:ion-yield}
\end{figure}

\begin{figure}[]
  \centering
  \includegraphics[width=0.4\textwidth]{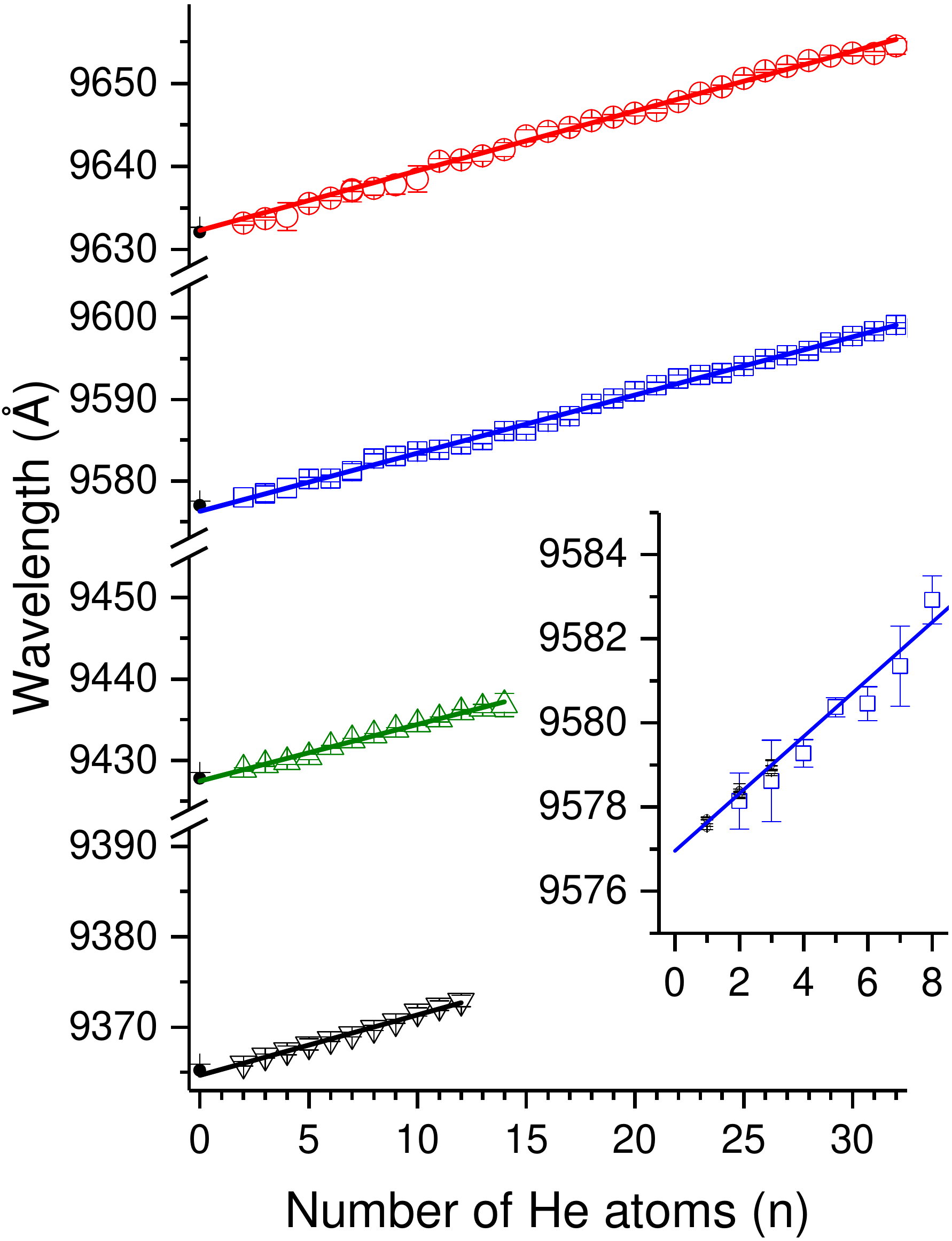}
  \caption{Absorption wavelength of \ce{He_$n$}\ce{C60+} as a function of the number of \ce{He} atoms attached. The four graphs (red, blue, green and black) correspond to bands 1, 2, 3, and 4. The inset shows how our values add to the data presented in \citet{Campbell2016b}. The values are given in air.}
  \label{fig:absorption-wavelength}
\end{figure}

\section*{Results and discussion} \label{sec:results}

Size dependent wavelength shifts as visible from the mass spectra in Figure 1 are summarized for four \ce{He_$n$}\ce{C60+} bands (bands 1, 2, 3, and 4) in Figure 2. Band 5 has not been investigated because of the low signal intensity. In this diagram the absorption wavelengths for sequential $n$-values are given, with $n$ varying from 2$-$32, 2$-$32, 2$-$14, and 2$-$12 for bands 1, 2, 3, and 4, respectively. Each additional \ce{He} atom causes a small and characteristic shift of $\sim$~0.7~\AA. This effect is found for all four \ce{C60+} bands and is fully determined by the way a solid layer of helium atom forms around the \ce{C60+} that can be considered an ionic impurity in bulk superfluid helium. From the experiment and supporting theory \citep[see][]{Kuhn2016b} it is confirmed that the first shell around the \ce{C60+} is filled by 32 \ce{He} atoms; 20 \ce{He} atoms above hexagons, and 12 \ce{He} atoms above pentagons. The binding energy of \ce{He} to \ce{C60+} onto a hexagonal and a pentagonal face is not identical (17 vs. 15~meV according to path integral calculations), so one would expect that a \ce{He} attached to a hexagonal site leads to a somewhat stronger redshift than to a pentagonal site. But as we do not find any kink at $n=12$ or 20, it is very likely that all sizes are simply statistically occupied with no clear isomerization preference for specific sites. This will be the topic of a separate study.

At $n=32$, a clear change from a linear red shift to a non- or much less linear blue shift is also found. This reflects perturbations caused by the ongoing implementation of \ce{He} atoms in the first shell, delocalizing the already attached \ce{He} atoms. This solvation behavior has been investigated in detail by \citet{Kuhn2016b} and, as stated before, is not the topic of the present study. The goal here is to use up to 32 measurement points for the four stronger \ce{C60+} bands to extrapolate to the wavelength for which the bare \ce{C60+} ion is expected to absorb, as illustrated in Figure 2. The number of data points for the two weaker bands is smaller than for bands 1 and 2, but still substantially larger than the number of measurement points obtained in the \ce{C60+}$-$\ce{He_$n$} ion trap tagging experiment. Error bars indicate the position uncertainty of the Lorentzian fit of each resonance, which takes the standard deviation of the ion yield into account. Typical standard deviations amount to 0.1~\AA. The absolute laser wavelength has been determined using a wavelength meter (HighFinesse WS-7) with an estimated systematic deviation of $s=\pm$~0.1~\AA. 

In the first two columns of Table~1 with laboratory data, we compare the transition wavelengths of \citet{Kuhn2016b} with the values derived in the present study. The wavelengths are given both in vacuum and air, using the \citet{Morton2000} IAU standard conversion. It should be noted that the extrapolation shown here is based on the physically maximum number of data points accessible -- beyond 32 \ce{He} atoms, the \ce{C60+} fullerene cations complex along a different mechanism and the resulting shifts cannot be used anymore \citep[see][]{Kuhn2016a}. The inset of Figure 2 illustrates the problem that exists when extrapolating the rest wavelengths based on a smaller sample issue; a small deviation in one of the measuring points can already affect the final rest wavelength substantially. This becomes a problem when the deviation is of the order of the astronomical accuracy with which DIB peak positions can be determined. This may explain why the initial values reported by \citet{Campbell2015} were corrected recently, shifting them by about $0.5-0.7$~\AA\ \citep{Campbell2016a}. These values overlap and add a few wavelength points to the lower end of our data set. As the estimated uncertainties are rather similar, these can be included in a more complete fit without additional weighting, resulting in peak positions at 9634.7(1), 9579.5(1), 9430.1(1), and 9367.5(1)~\AA\ (in vacuum). The corresponding wavelengths in air are also listed in Table~1. These values slightly deviate (between 0.1 up to 0.5~\AA) from those previously reported in \citet{Kuhn2016b}. This is partly due to a re-evaluation of our data with better corrections for contaminations, such as pristine \ce{He} clusters. The final values are within 0.3~\AA\ from the values reported by \citet{Campbell2016b}. Our overall precision has increased because of the use of more data points and an additional measurement series.

In Table~1, all values reported so far are summarized and compared with the astronomical values. This is also visualized in Figure~3, showing the artificial DIB profiles for bands 1, 2, 3, and 4 in normalized intensities. The shaded area shows the astronomical FWHMs of the bands involved.

\begin{figure}[ht!]
  \centering
  \includegraphics[width=0.4\textwidth]{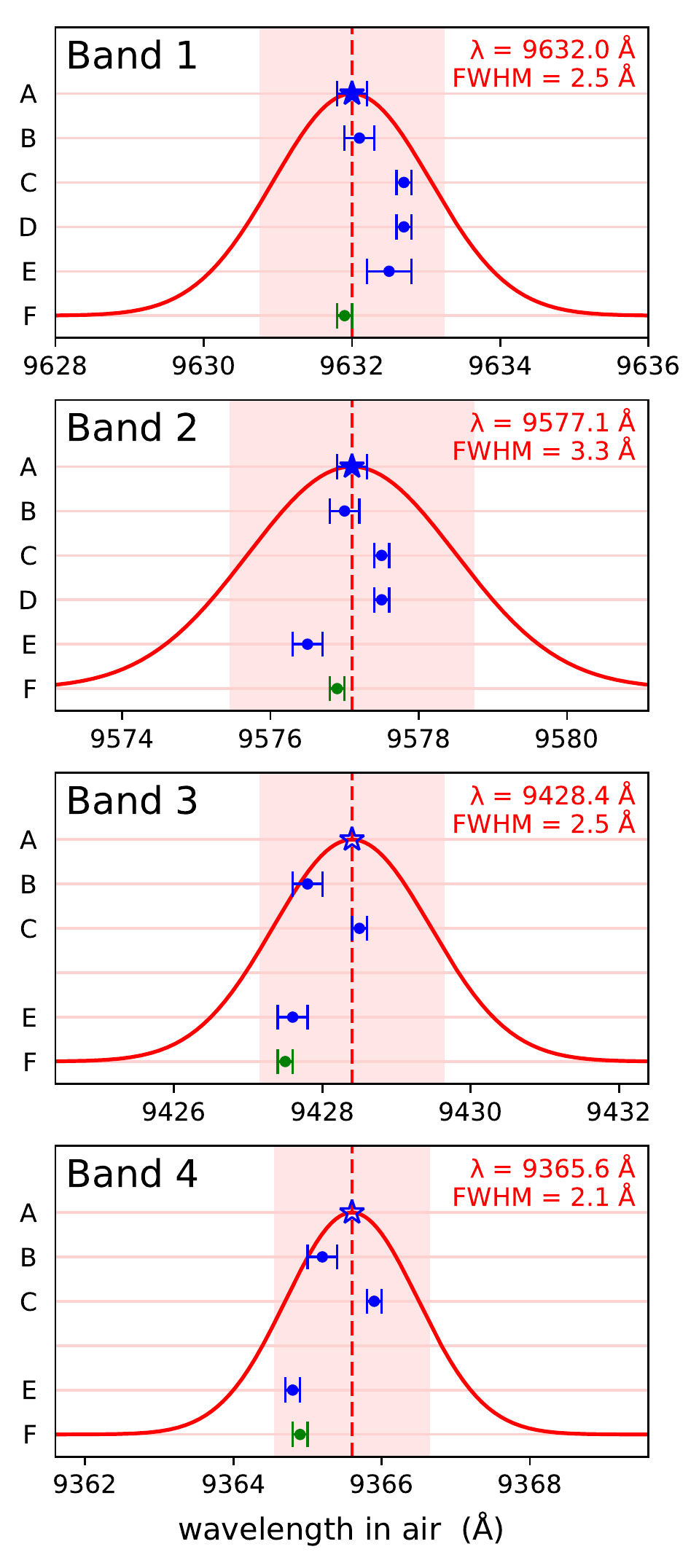}
  \caption{A comparison of the astronomical DIB data (A) with all laboratory data, obtained in the \ce{He}-tagged ion trap experiments by \citet{Campbell2016b,Campbell2016a,Campbell2015} (B, C, D), in the \ce{He} droplet experiment by \citet{Kuhn2016b} (E), and in the present work (F). The red curves are the (normalized) astronomical profiles and the red shaded zone shows the astronomical FWHM as reported in \citet{Cox2014b} or \citet{Walker2016}. All wavelengths are in air.}
  \label{fig:comparison}
\end{figure}

In the most recent paper \citet{Walker2016} used the \citet{Campbell2016a} data to conclude that DIB features derived from the Canada-France-Hawaii Telescope data do fit even better with these adapted laboratory values. As stated before, bands 1 and 2 have been recorded in a number of independent astronomical studies; for bands 3, 4, and 5 this is not the case yet and the astronomical peak positions were guided by the work presented in \citet{Walker2015,Walker2016}.

\begin{table*}[!]
\centering
\caption{Wavelength positions [\AA] of \ce{C60+} absorption lines obtained in the laboratory and the corresponding DIBs}
\label{tab:table1}

\resizebox{\textwidth}{!}{%
\begin{tabular}{@{}llccccccc@{}}
\toprule[1pt]
 && \multicolumn{5}{c}{Laboratory data} &  & \multicolumn{1}{c}{DIB data} \\ \cmidrule(lr){3-7} \cmidrule(l){9-9} 
 && \multicolumn{1}{c}{This work} & \citet{Kuhn2016b} & \citet{Campbell2015} & \citet{Campbell2016a} & \citet{Campbell2016b} &  & \begin{tabular}[c]{@{}c@{}}$^a$\citet{Cox2014b}\\ $^b$\citet{Walker2016}\end{tabular}\\
\midrule[1pt]
Band 1 & \begin{tabular}[c]{@{}l@{}}Vacuum\\Air\end{tabular} & \begin{tabular}[c]{@{}c@{}}9634.7(1)\\ 9631.9(1)\end{tabular} & \begin{tabular}[c]{@{}c@{}}9635.2(3)\\ 9632.5(3)\end{tabular} & \begin{tabular}[c]{@{}c@{}}\newline\\ 9632.7(1)\end{tabular} & \begin{tabular}[c]{@{}c@{}}\newline\\ 9632.7(1)\end{tabular} & \begin{tabular}[c]{@{}c@{}}\newline\\ 9632.1(2)\end{tabular} &  & \begin{tabular}[c]{@{}c@{}}\newline\\ 9632.0(2)$^a$\end{tabular} \\
 &  &  &  &  &  &  &  \\
Band 2 & \begin{tabular}[c]{@{}l@{}}Vacuum\\Air\end{tabular} & \begin{tabular}[c]{@{}c@{}}9579.5(1)\\ 9576.9(1)\end{tabular} & \begin{tabular}[c]{@{}c@{}}9579.1(2)\\ 9576.5(2)\end{tabular} & \begin{tabular}[c]{@{}c@{}}\newline\\ 9577.5(1)\end{tabular} & \begin{tabular}[c]{@{}c@{}}\newline\\ 9577.5(1)\end{tabular} & \begin{tabular}[c]{@{}c@{}}\newline\\ 9577.0(2)\end{tabular} &  & \begin{tabular}[c]{@{}c@{}}\newline\\ 9577.1(2)$^a$\end{tabular} \\
 &  &  &  &  &  &  &  \\
Band 3 & \begin{tabular}[c]{@{}l@{}}Vacuum\\Air\end{tabular} & \begin{tabular}[c]{@{}c@{}}9430.1(1)\\ 9427.5(1)\end{tabular} & \begin{tabular}[c]{@{}c@{}}9430.2(2)\\ 9427.6(2)\end{tabular} & \begin{tabular}[c]{@{}l@{}} \\$\cdots$\end{tabular} & \begin{tabular}[c]{@{}c@{}}\newline\\ 9428.5(1)\end{tabular} & \begin{tabular}[c]{@{}c@{}}\newline\\ 9427.8(2)\end{tabular} &  & \begin{tabular}[c]{@{}c@{}}\newline\\ 9428.4$^b$\end{tabular} \\
 &  &  &  &  &  &  &  \\
Band 4 & \begin{tabular}[c]{@{}l@{}}Vacuum\\Air\end{tabular} & \begin{tabular}[c]{@{}c@{}}9367.5(1)\\ 9364.9(1)\end{tabular} & \begin{tabular}[c]{@{}c@{}}9367.4(1)\\ 9364.8(1)\end{tabular} & \begin{tabular}[c]{@{}l@{}} \\$\cdots$\end{tabular} & \begin{tabular}[c]{@{}c@{}}\newline\\ 9365.9(1)\end{tabular} & \begin{tabular}[c]{@{}c@{}}\newline\\ 9365.2(2)\end{tabular} &  & \begin{tabular}[c]{@{}c@{}}\newline\\ 9365.6$^b$\end{tabular} \\
 &  &  &  &  &  &  &  \\
Band 5 & \begin{tabular}[c]{@{}l@{}}Vacuum\\Air\end{tabular} & \begin{tabular}[c]{@{}l@{}}$\cdots$\\$\cdots$\end{tabular} & \begin{tabular}[c]{@{}l@{}}$\cdots$\\$\cdots$\end{tabular} & \begin{tabular}[c]{@{}l@{}} \\$\cdots$\end{tabular} & \begin{tabular}[c]{@{}c@{}}\newline\\ 9349.1(1)\end{tabular} & \begin{tabular}[c]{@{}c@{}}\newline\\ 9348.4(2)\end{tabular} &  & \begin{tabular}[c]{@{}c@{}}\newline\\ 9348.4$^b$\end{tabular} \\
\bottomrule[1pt]
\\
\multicolumn{9}{l}{Note: The listed DIB values are taken from the catalogued DIB lists \citep{Cox2014b,Walker2015,Walker2016}. The uncertainty in the values from this}\\
\multicolumn{9}{l}{work reflects the statistical error.}
\\
\end{tabular}
}
\end{table*}

The \ce{C60+} rest wavelengths derived from the \ce{He} droplet experiments presented here confirm the initial conclusions made in \citet{Campbell2015}. Within the relatively small error margins of the individual experiments the overlap is convincing, typically within 2$\sigma$ or better. Even though no pure \ce{C60+} ions were measured directly in the ion trap study, nor in our \ce{He} droplet experiment, it is fair to conclude that the extrapolations provide data accurate enough (within 0.2~\AA) to compare with the astronomical spectra. This accuracy is smaller than the typical bandwidths found in astronomical spectra, as shown in Figure~3. The experimental bandwidths reported by Campbell et al. for the small \ce{C60+}$-$\ce{He_${1-3}$} clusters are of the order of $2.2-2.5$~\AA. This value may have an isomeric origin; it makes a difference, here, whether a \ce{He} atom is situated above a hexagon or a pentagon site, i.e., one \ce{He} atom already can result in two energy values. For the bigger clusters, in our experiment, the larger number of different isomeric forms will contribute to the overall bandwidths. The only exception is \ce{He32}\ce{C60+} species for which only one isomeric form exists and the experimental bandwidth determined here amounts to 2.4~\AA, comparable to the values derived by Campbell and coworkers and very similar to the astronomical values that are listed in Fig. 3. Effects of Doppler broadening can be safely neglected.

Apart from an independent check of the rest wavelengths and bandwidth of \ce{C60+}, the \ce{He} droplet measurements presented here also provide information on the intensity ratios of the \ce{C60+} bands. In the ion tag experiments the intensity ratio of bands 1 to 4 is 0.8 : 1.0 : 0.3 : 0.2. \citet{Galazutdinov2017} concluded that the astronomical intensity ratios of bands 1 and 2, determined for a large number of lines-of-sight, do not behave like 0.8 : 1.0, and actually seem to lack any correlation at all. This has been used as an argument against \ce{C60+} as the carrier of these bands. Galazutdinov and coworkers used theoretical modeling of contaminating \ce{Mg\ \textsc{ii}} stellar lines to derive their conclusions. This approach was questioned by \citet{Walker2017} who concluded equivalent width ratios fully in line with the laboratory values using a different method. A discussion about the astronomical data interpretation is beyond the scope of the present work. However, it is possible that intensity ratios derived from a complexed fullerene ion do not reflect those of the bare \ce{C60+} cation. For this reason, we have investigated the intensity ratio for bands 1 and 2 for $n$-values up to 100. These experiments turned out to be hard, but show that the ratio is not fully constant and can (slowly) vary with sequential $n$-values. The effect is particularly visible for higher $n$-values; for low $n$-values the error margins are too large to derive conclusive statements. It is likely, however, that the intensity ratio for the bare \ce{C60+} bands will not differ much from that derived experimentally for lower $n$-values. We are currently in the process of remeasuring the relative intensities of bands 1 to 4 for $n$-values below $n=10$.

A logical continuation at this stage is an independent observational check to investigate to which extent also the weaker DIBs can be found at the experimentally determined wavelengths. Heavy telluric pollution in this wavelength region, particularly by water features, complicates the astronomical data interpretation. As stated earlier, a first attempt to use the Hubble Space Telescope for DIB research in the 1~$\mu$m region has been successful \citep{Cordiner2017}, but signals for bands 3, 4, and 5, if any, were found to be at the noise level and a follow up study for different target stars is needed for an unambiguous conclusion.

Both \ce{C60} \citep{Cami2010,Sellgren2010} and \ce{C60+} \citep{Berne2013} have been detected in other environments in space through their mid-infrared emission spectra. Given the relatively low ionization potential of \ce{C60} \citep[$\sim$~7.6~eV,][]{DeVries1992}, one would expect \ce{C60+} (or possibly even \ce{C60}$^{2+}$) to be observable as well. Clearly, a positive identification of \ce{C60+} in translucent interstellar clouds would be important from an astrochemical point of view and the combined work by Campbell and Walker strongly hint in this direction. Up to now only \ce{C3} \citep{Haffner1995,Maier2001,Schmidt2014} has been identified as one of the few smaller and pure carbon species in diffuse clouds. Attempts to search for \ce{C4} and \ce{C5} resulted in upper limits \citep{Maier2004}. It is obvious that \ce{C60+} will not form from merging twenty \ce{C3} units and a better idea of the role of polyatomic carbon species is needed to understand the processes at play. Possibly, \ce{C60} does not form bottom-up, but in top-down processes, i.e., as reaction product from a larger precursor. In a recent study by \citet{Zhen2014} it was experimentally shown that fullerenes can form upon photo-dissociation of GRAND PAHs, very large polycyclic aromatic hydrocarbons (PAHs), fully in agreement with predictions by \citet{Berne2012} and \citet{Berne2015}. This would link DIBs to PAHs, a hypothesis that has been tested in the past \citep[see][for an overview]{Cox2014a}, but laboratory spectra of commercially available PAHs and their cations did not match with reported DIB features \citep{Salama2011}. Instead, it may be possible that large PAH derivatives or other, smaller and more stable carbon cages are involved. Such species have been hard to produce in large abundances and under mass resolved and fully controlled laboratory conditions. The method presented here, however, has the sensitivity to exactly provide spectra for such species.

In conclusion, our work shows that the available \ce{C60+} laboratory rest wavelengths and derived bandwidths can be used for comparison with astronomical data. Band intensity ratios do not seem to vary strongly between subsequent $n$-values.  An  unambiguous identification of \ce{C60+} as a DIB carrier is awaiting a pollution-free spectrum, clearly exhibiting all five \ce{C60+} bands with equivalent width ratios and bandwidths in agreement with the available laboratory data.

\balance

\subsection*{Acknowledgments}
This study was supported by the Fonds zur F{\"o}rderung der Wissenschaftlichen Forschung (FWF) projects P26635, W1259, and I978-N20, Deutsche Forschungsgemeinschaft (DFG) project I978-N20, the European COST Action CM1204 XLIC, and the European Research Council under ERC Grant Agreement Number 279898. Laboratory astrophysics in Leiden is supported by the Netherlands Research School for Astronomy (NOVA) and NWO (Netherlands Organisation for Scientific Research) through the Dutch astrochemistry network and a VICI grant.

\centering\rule[-5mm]{3cm}{1pt}



\bibliographystyle{elsarticle-harvard} 

\vspace{20pt}

\end{document}